\documentclass[floatfix,twocolumn,showpacs,preprintnumbers,amsmath,amssymb,prb]{revtex4}

\usepackage{graphicx}
\usepackage{dcolumn}
\usepackage{bm}
\usepackage{psfig}

\newcommand{\beq}{\begin{equation}}
\newcommand{\eeq}{\end{equation}}
\newcommand{\sgn}{{\rm sgn}}
\newcommand{\kv}{{\bf k}}
\newcommand{\xv}{{\bf x}}
\newcommand{\Qv}{{\bf Q}}
\newcommand{\qv}{{\bf q}}
\newcommand{\pv}{{\bf p}}

\newcommand{\beqa}{\begin{eqnarray}}
\newcommand{\eeqa}{\end{eqnarray}}
\newcommand{\qav}[1]{\langle #1 \rangle}
\newcommand{\refe}[1]{(\ref{#1})}
\newcommand{\refE}[1]{Eq.~(\ref{#1})}
\newcommand{\refF}[1]{Fig.~\ref{#1}}
\newcommand{\Gc}{{\cal G}}
\newcommand{\Fc}{{\cal F}}

\newcommand{\Vc}{{\cal V}}
\newcommand{\Tc}{{\cal T}}

\newcommand{\rem}[1]{}

\begin{document}


\title{Superconductivity with hard-core repulsion:\\
BCS-Bose crossover and s-/d-wave competition
}

\author{F.~Pistolesi}\email{Fabio.Pistolesi@grenoble.cnrs.fr}
 \altaffiliation[present address: ]{
Laboratoire de Physique et Mod\'elisation des Milieux Condens\'es,
Centre National de la Recherche Scientifique, 
BP~166, 38042 Grenoble Cedex 9, France
}
\affiliation{
European Synchrotron Radiation Facility\\
BP~220, 38043 Grenoble Cedex 9, France
}

\author{Ph.~Nozi\`eres}
\affiliation{
Laboratoire d'Etude des Propri\'{e}t\'{e}s Electroniques des Solides\\
Centre National de la Recherche Scientifique\\
BP~166, 38042 Grenoble Cedex 9, France}

\date{\today}

\begin{abstract}
We consider fermions on a 2D lattice interacting repulsively on the
same site and attractively on the nearest neighbor sites.
The model is relevant, for instance, to study the competition between
antiferromagnetism and superconductivity in a Kondo lattice.
We first solve the two-body problem to show that in the dilute and
strong coupling limit the $s$-wave Bose condensed state is always
the ground state.
We then consider the many-body problem and treat it at mean-field
level by solving exactly the usual gap equation.
This guarantees that the superconducting wave-function correctly
vanishes when the two fermions (with antiparallel spin) sit on the
same site.
This fact has important consequences on the superconducting state 
that are somewhat unusual. 
In particular this implies a radial node-line for the gap function.
When a next neighbor hopping $t'$ is present we find
that the $s$-wave state may develop nodes on the Fermi surface.

\end{abstract}

\pacs{
{74.20.-z}, 
{74.20.Fg}, 
{74.20.Mn} 
\hskip 2cm {\bf Published:} Phys. Rev. B {\bf 66} 054501 (2002)
}

\maketitle

\section{\label{introduction} Introduction}

Superconductivity is possible when carriers attract.
The attraction in metals is normally due to the phonon coupling. 
The strong coulomb repulsion is normally well screened in metals
and under some conditions it can be overcome by the attraction
leading to superconductivity.
A locally attractive (non-retarded) model, such as the BCS one, can thus
give a good qualitative description of the superconducting state since
the correlation length $\xi$ is much larger than both  the screening
length and the interparticle density.

This is not the case when the carriers are non conventional.
A relevant example is the single-band Kondo lattice with a RKKY
interaction.
For large Kondo coupling the ground state of this system is formed by
local singlets: it is thus insulating at half-filling.
At small doping the carriers are bachelor spins: they feel a
hard core repulsion, since they cannot sit on the same site at the 
same time. 
The RKKY antiferromagnetic exchange interaction  gives an attraction
among holes in the singlet channel, and thus it could 
lead to a superconducting state \cite{Andrei,nozieres}.
But in this case the hard-core repulsion cannot be
neglected in the description of the $s$-wave superconductivity.
More details on the Kondo lattice model are discussed in 
Reference \onlinecite{nozieres}.

In this paper we consider a simple model where the role of the
hard-core repulsion can be analyzed in a transparent way.
For simplicity and relevance to physical systems we 
consider a two dimensional square lattice with hopping and 
attractive interaction among nearest neighbor sites. 
Different range for the interaction and the hopping produces
interesting results, that will be analyzed.
Similar models have been studied by different authors.
Ohkawa and Fukuyama proposed an ansatz to solve the mean-field
equations for superconductivity with local repulsion
in Kondo systems \cite{Fukuyama}.  
Later Micnas, Ranninger, and Robaszkiewicz 
\cite{Ranninger} and Bastide, Lacroix, and Rosa Simoes
\cite{BastideLacroix}, applied that technique to calculate 
the superconducting critical temperature.
Similar equations in a different context were considered 
by Aligia and coworkers.\cite{Aligia}

We will reconsider the problem at zero temperature 
and investigate the competition among
different symmetries of the order parameter as one spans the crossover
from BCS superconductivity to Bose condensation 
\cite{Legget,NSR,ranninger,RanderiaPRL,Haussmann,PistolesiStrinati,%
zwerger,Levin,Hertog,andrenacci}. 
In particular we will clarify an apparent disagreement with 
the prediction of Randeria \cite{Randeria2D} on the presence of 
bound states and superconducting instabilities in 2D systems.
We will also discuss the unusual nature of the superconducting state:
the main peculiarity is the presence of a radial line of
nodes in $k$ space for the gap function $\Delta(\kv)$.

The plan of the paper is the following.
In Section \ref{model} we define in detail the model we are
considering.
In Section \ref{twobody} we solve the two body problem and we 
discuss the relevance to the Bose limit.
In Section \ref{MeanField} we set up the mean field equations 
and find the resulting phase diagram for the ground state.
The competition between $s$- and $d$-wave superconductivity is
analyzed in Section \ref{dwave}.
In Section \ref{tprimo} the effect of a mismatch between the 
hopping and the interaction range is considered by introducing a
hopping $t'$ among next-neighbor sites. 

\section{The model}
\label{model}

We consider fermions hopping on a 2D square lattice of spacing $a$
with nearest-neighbor sites hopping matrix elements $t$.
(In some cases we will consider also next-neighbor hopping $t'$.)
The fermions interact with a local repulsive interaction parameterized
by $U$, that we will set to infinity.
They furthermore feel an attraction $V$, which we assume to act only
between nearest neighbors (in a real RKKY example that would not be the case) 
with opposite spins. 
The relevant parameters at zero temperature are only $V/t$ and 
the filling of the band $0<n<2$.
The spin does not play an essential role. 
Since we are interested in the singlet state, we can disregard
the parallel-spin part of the interaction, the mean field equations
would not be modified by that.
The model Hamiltonian is thus:
%
\beqa
	H &=& 
	\sum_{\kv \sigma} (t_\kv -\mu) 
	c^\dag_{\kv\sigma} c^{\vphantom\dag}_{\kv\sigma}
	+
	U \sum_i (n_{i\uparrow}-{1\over 2}) (n_{i\downarrow}-{1\over 2}) 
	\nonumber\\
	&&-V \sum_{i, \delta} 
	(n_{i\uparrow}-{1\over 2}) (n_{i+\delta\downarrow}-{1\over 2}) 
	\label{Ham1}
\eeqa
where the vector $\delta$ spans the four nearest neighbor sites
of the square lattice, $\sigma$ is the spin projection in a chosen
direction and $t_\kv = -2t [\cos(k_x a)+\cos(k_y a)]$
is the single-particle dispersion relation.

The Hamiltonian in the form of Eq.~\refe{Ham1} is particle-hole
symmetric:  for half-filling $\mu=0$. 
For convenience it can also be written in a non p-h
symmetric, but shorter form:
\beq
	H = 
	\sum_{\kv \sigma} (t_\kv -\mu_o) 
	c^\dag_{\kv\sigma} c^{\vphantom\dag}_{\kv\sigma}
	+
	U \sum_i  n_{i\uparrow} n_{i\downarrow}
	-V \sum_{i, \delta} n_{i\uparrow} n_{i+\delta\downarrow}
	\,,
	\label{Ham2}
\eeq
where
$
	\mu_o  =  \mu + (U-4V)/2
	\label{originalshift}
$.
The two forms are equivalent but from a constant.
The interaction term can be written in Fourier space as usual: 
\beq
	\sum_{\kv_1+\kv_2=\kv_3+\kv_4} V(\kv_2-\kv_3)
	c^\dag_{\kv_1\uparrow} c^\dag_{\kv_2\downarrow} 
	c^{\vphantom \dag}_{\kv_3\downarrow} 
	c^{\vphantom \dag}_{\kv_4\uparrow} 
\eeq
with 
$
	V(\kv) = U-V w(\kv)
$
and 
$
	w(\kv) = \sum_\delta e^{i \kv \delta} =  
	2 [\cos(k_x a)+\cos(k_y a)]
$.

In the following we investigate the superconducting order in 
the ground state of the Hamiltonian \refe{Ham1}.
In order to treat the strong repulsion correctly we 
consider first the two body problem, that is relevant in the 
dilute limit.

\section{Two-body problem}
\label{twobody}

If the attractive interaction is strong enough to bind a pair
we expect to reach the Bose condensed limit for a low density 
of carriers.
It is thus crucial to know whether a bound state is present or 
not. 
In our case the hard-core repulsion will fight against an 
$s$-wave bound state, while it will not affect the 
$d$-wave state.
In the dilute limit the choice between $s$- and $d$-wave 
is determined simply by the lowest 2-body bound state, 
if present. 

We thus begin our investigation by solving the two body problem: 
we find the threshold for the appearance of bound states.
The two-body problem  is exactly solvable by diagonalization
of a 5 by 5 matrix, corresponding to the central and 4 neighbor sites
involved in the equations of motions.
This can be done directly in real space, but we will proceed in
momentum space in order to use the same approach for the mean field
equations in the next Section.
In $k$-space the simplification comes from the separability 
of the interaction potential.
As a matter of fact,  $V_{\kv \kv'}=V(\kv-\kv')$, defined above 
can be written in a separable form as follows:
\beq
	V_{\kv \kv'} 
	=
	\sum_{\alpha\beta} 
	  w_\alpha (\kv) \, \Vc_{\alpha\beta}\, w_\beta(\kv')
\eeq
where
\beq	
	\left\{
	\begin{array}{rcl}
	w_0(\kv) &=& 1\\
	w_1(\kv) &=& [\cos(k_x a)+\cos(k_y a)]/\sqrt{2}\\
	w_2(\kv) &=& [\cos(k_x a)-\cos(k_y a)]/\sqrt{2}\\
	w_3(\kv) &=& \sin(k_x a)\\
	w_4(\kv) &=& \sin(k_y a) 
	\end{array}
	\right.
	\,,
\eeq
$\Vc_{\alpha\beta}= v_\alpha \delta_{\alpha \beta}$, 
with $v_{0}=U$  and $v_\alpha=-2V$ for $\alpha \neq 0$.
We have chosen a symmetrized and anti-symmetrized combination
of cosines in order to exploit the symmetry of the lattice.

The Green's function  for the two-body problem 
on the basis 
$
	|\qv \Qv\rangle = 
	c^\dag_{\qv+\Qv/2\uparrow} 
	c^\dag_{-\qv+\Qv/2\downarrow}
	|0\rangle
$
is diagonal in $\Qv$. We consider only the case $\Qv=0$ 
\cite{Qdep}.
The corresponding equation of motion is thus that of a particle
in an external potential:
\beqa
	(\omega-2t_{\kv})\Gc_{\kv \pv}(\omega)
	- \sum_\qv V_{\kv, \qv}\Gc_{\qv \pv}(\omega)
	= \delta_{\kv \pv}
	\,.
\eeqa

Since $V$ is separable the above integral equation reduces 
to an algebraic one for 5 parameters.
The substitution of the separable form of $V$ gives:
\beq
	(\omega-2 t_\kv) 
	\Gc_{\kv \pv}(\omega) -
	\sum_\alpha  w_\alpha(\kv) A_{\alpha \pv}(\omega)
	=
	\delta_{\kv \pv}
	\,,
\eeq
where 
\beq
	A_{\alpha \pv}(\omega)
	= 
	v_\alpha \sum_\qv  w_\alpha(\qv) \Gc_{\qv \pv}(\omega)
	\,. \label{Adef}
\eeq
The Green's function is thus:
\beq
	 \Gc_{\kv \pv}(\omega) =
	\Gc^o_{\kv}(\omega)
	\left[
	\delta_{\kv \pv} +
	\sum_\alpha  w_\alpha(\kv) A_{\alpha\pv}(\omega)
	\right]
	\,, \label{GcA}
\eeq
where
$
	\Gc^o_{\kv}(\omega)
	=
	(\omega-2t_\kv)^{-1}
$. 
Substituting \refe{GcA} into \refe{Adef}
we obtain the following closure equation for $A_{\alpha \pv}$:
\beq
	A_{\alpha \pv}(\omega) 
	=  
	v_\alpha  w_\alpha(\pv) \Gc^o_\pv(\omega)
	+ 
	v_\alpha \sum_{\beta}   
	B_{\alpha \beta}(\omega)  
	A_{\beta \pv}(\omega)
	\,, \label{selfequations}
\eeq
where 
\beq
	B_{\alpha \beta}(\omega) = 
	\sum_\qv 
	w_\alpha(\qv) w_{\beta}(\qv)  \Gc_\qv^o(\omega) 
	\,.
\eeq
The set of linear equations  \refe{selfequations} 
for $A$ can be solved by inversion of the 5x5 matrix,
$
	1-\Vc\,B(\omega)
$ 
\beq
	A_{\alpha \pv } (\omega)
	= 
	\sum_{\beta} \Tc_{\alpha \beta}(\omega)
	w_\beta(\pv) \Gc^o_{\pv}(\omega) 	
\eeq
where we have defined $\Tc(\omega)=[1-\Vc B(\omega)]^{-1}\Vc$.
The final expression for the Green's function is:
\beq
	\Gc_{\kv \pv}(\omega)   = 	
	\Gc^o_{\kv}(\omega)
	\left[
	\delta_{\kv \pv} + 	
	\Gc_\pv^o(\omega)
	\sum_{\alpha\beta} 
	\Tc_{\alpha \beta}(\omega)
	w_\alpha(\kv) 
	w_\beta(\pv) 
	\right]
	\,.
\eeq

Note that $\Tc$ is the usual T-matrix of scattering theory:
\beq
	T_{\kv \pv} = \sum_{\alpha \beta} 
	\Tc_{\alpha \beta} w_\alpha(\kv) w_{\beta}(\pv)
	\,.
\eeq

The problem can now be solved for any value of $U$, but since 
we are interested in the $U=\infty$ limit it is convenient to
write $\Tc$ as:
\beq
	\Tc = (\Vc^{-1}-B)^{-1}
\eeq
where  $(\Vc^{-1})_{00}=1/U=0$. 
In this way we eliminate $U$ from the outset.
One can readily verify that for $U=\infty$ the resulting local Green
function $\Gc(\xv_i=0)$ actually vanishes.
As it should, the hard-core repulsion prevents
two fermions with opposite spin projection to sit on the same site.

\begin{figure}
\psfig{file=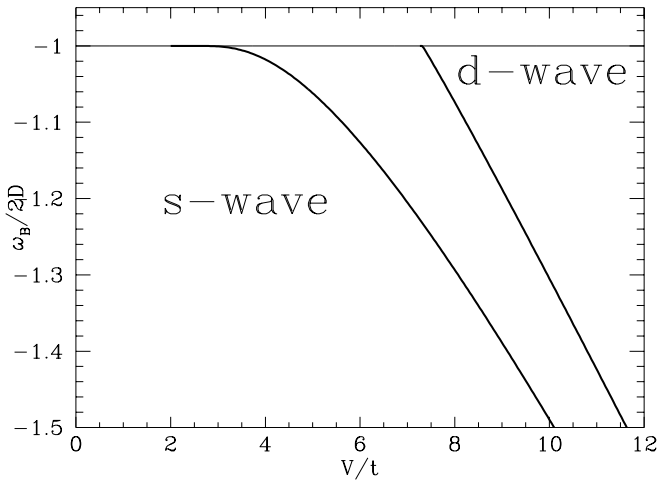,width=9cm,angle=-90}
\caption{Two-body $s$-wave and $d$-wave bound state energy as a function of $V/t$
and measured in terms of the band width $2D=8t$. 
In both cases we find a threshold for the appearance of the bound 
state.
}
\label{figbound}
\end{figure}

Let us come back to the explicit evaluation of $\Tc$ in two dimensions.
The $B$ matrix is in block form, $\Gc^o_\qv$ is in fact even under
both $k_x\rightarrow -k_x$ and $k_y\rightarrow -k_y$.
$w_2$ is odd under $k_x \leftrightarrow k_y$, and $w_3$ and $w_4$ are odd under
$k_x\rightarrow -k_x$ and $k_y\rightarrow -k_y$ respectively.
These symmetries leave few non vanishing elements and split the 
matrix in block form, corresponding to the irreducible representations 
of the symmetry group of the lattice:
\beq
	B = 
	\left( \begin{array}{ccccc} 
	B_{00} & B_{01} & 0      & 0  & 0      \\
	B_{01} & B_{11} & 0      & 0  & 0      \\
	0      & 0      & B_{22} & 0 & 0       \\
	0      & 0      & 0      & B_{33} & 0  \\
	0      & 0      & 0      & 0  & B_{33} \\
	\end{array}
	\right) 
	\,.
\eeq
The bound states are given by the solution of the equation
$\det(\Vc^{-1}-B)=0$.
We obtain thus the following equations:
\beqa
	{1\over 2V} &=& {B_{01}^2 \over B_{00}} -B_{11}  
	\,,
	\label{Swave}
	\\
	-{1\over 2V} &=& B_{22} 
	= 
	{1\over 2} \sum_{\kv} {
	[\cos(k_x a)-\cos(k_y a)]^2
	\over \omega -2t_\kv}
	\,,
	\label{Dwave}
	\\
	-{1\over 2V} &=& B_{33} 
	= \sum_{\kv} {\sin^2(k_x a) \over \omega -2t_\kv}
	\label{Pwave}
	\,.
\eeqa
Eqs.~\refe{Swave}, \refe{Dwave}, and \refe{Pwave} refer to 
$s$-, $d$-, and $p$-wave bound states, respectively.
An additional $s$-wave bound state has been eliminated by 
setting $U=\infty$. 
Note that $U$ does not affect the $p$- and $d$-solutions,
since the wavefunction vanishes at the origin.
In contrast, the $s$-wave bound state is strongly affected by $U$.
As a matter of fact, $B_{00}(\omega)$, $B_{01}(\omega)$, and
$B_{11}(\omega)$ all diverge logarithmically for $\omega\rightarrow
2D=8t$, with $2D$ the single particle band width.
Using Eqs. \refe{ee1} and \refe{ee2} below one can verify that these divergences
disappears in Eq.~\refe{Swave}, due to mutual cancellation:
the
$s$-wave bound state is present only for $V$ larger than a threshold
$V_t^s$,
while for a purely attractive potential binding always occurs 
in two-dimensions.

It is not difficult to prove the following exact relations
among the different $B$'s by summing and subtracting $\omega$ and 
cosine terms in the numerator:
\beqa
	&&\omega B_{00}(\omega) + 4\sqrt 2\, t B_{01}(\omega) = 1  
	\,, \label{ee1}
	\\
	&& 
	\omega B_{01}(\omega)+4 \sqrt2\, t B_{11}(\omega) = 0
	\,. \label{ee2}
\eeqa
Substituting Eqs.~\refe{ee1} and \refe{ee2} into Eq.~\refe{Swave}
the $s$-wave bound state equation becomes:
\beq
	{2t\over V} 
	= -{\omega \over 8 t} + {1 \over 8 t\,B_{00}(\omega_B)}
	\qquad 
	s{\rm -wave}
	\label{finalswave}
	\,,
\eeq
where explicitly
$
	B_{00}(\omega) = 
	\sum_{\kv} 
	(\omega -2t_\kv)^{-1}
$.
At the threshold $V_t^s$ the bound state energy lies at the
band bottom $\omega=-8t$ where $B_{00}$ diverges: hence
the threshold $V^s_t$ for a $s$-wave bound state is simply
$V=2t$.
The $s$-wave solution appears non analytically at the threshold:
$\omega_B(V) \approx -2D - 4t \exp\{-\pi V_t^s/(V-V^s_t)\}$.
For values of $V>V_t^s$ Eq.~\refe{finalswave} can be solved numerically.
We plot the result in Figure \ref{figbound} together with the $d$-wave
bound state energy obtained by numerical solution of Eq.~\refe{Dwave}.
As it can be seen from the picture the $d$-wave threshold $V_c^d
\approx7.35 \,t$ is higher than the $s$-wave one, (despite 
the fact that the latter is affected by the hardcore).
This reflects the cost in kinetic energy (curvature of the wave functions)
of changing the sign as we go
around $ {\bf x} = 0$.

We thus conclude that for $V>V_t^s$ and for low density the 
ground state of the system is a Bose condensate of 
fermions pairs in a $s$-wave bound state. 
The Bose limit is safely achieved only if the bound
state energy, $ |\omega_B+2D| $, is much larger than the interparticle
distance energy scale, i.e. $ \epsilon_F \sim t\, n$.
Since $\omega_B+2D$ vanishes exponentially near $V_c$ 
the Bose region in the $n$-$V$ diagram rapidly shrinks as 
$V \rightarrow V_c$ (cf. Fig. \ref{FigVt} in the following).

This picture for dilute pairs is quite clear and it will not be modified
by the introduction of a small $t'$ (see following discussion).
The opposite BCS limit is on the other hand more subtle.

\section{Mean field equations for superconductivity}
\label{MeanField}

In order to study the onset of superconductivity for small
coupling and large density, we consider the usual BCS mean field
theory.

As in the two-body problem, the integral equation 
can be reduced to a simple algebraic equation. 
The $s$-wave solution is strongly modified by 
the hard-core interaction, and the order parameter has a 
$\kv$ dependence that comes from the mixing of the 
$w_0$ and $w_1$ terms, as in the two body problem.
This mixing is crucial in order to ensure the vanishing of 
the superconducting wavefunction at the origin.

The mean-field theory equations describing the
superconducting state of the Hamiltonian \refe{Ham1} are
obtained by the usual decoupling:
\beq
	H_{MF} = 
	\sum_{\kv \sigma} \xi_\kv
	c^\dag_{\kv\sigma} c^{\vphantom\dag}_{\kv\sigma}
	+
	\sum_{k} \left[
	\Delta_{\kv} c^\dag_{\kv\uparrow} c^{\dag}_{-\kv\downarrow}
	+ cc
	\right]
	\,.
\eeq
The minimization procedure gives the familiar equations for 
the ground state:
\beqa
	n &=& \sum_{\kv}
	\left[1-{\xi_\kv \over E_\kv}  \right]
	\,,
	\label{nequation}
	\\
	\Delta_\kv &=& -\sum_{\kv'} V(\kv-\kv') 
	{\Delta_{\kv'} \over 2 E_{\kv'}} 
	\,,
	\label{gapequation}
\eeqa
where  	
$
	\xi_\kv = t_\kv -\mu_o+ V(0) n/2
$, $
E_\kv =\sqrt{\xi_\kv^2+|\Delta_\kv|^2}$,
and for consistency we retained the Hartree term. 
Putting together the Hartree shift and the original shift
of the chemical potential we have
$
	 \xi_\kv = t_\kv -\mu',
$
with
$
	\mu' = \mu + {1\over 2} (U-4V)(n-1)
$.
(There is no Fock exchange term because of our interaction between
opposite spins only.)

We concentrate now on the gap equation \refe{gapequation}.
We exploit again the separability of the interaction to 
write:
\beq
	\Delta_\kv = - 
	\sum_\alpha v_\alpha w_\alpha(\kv) 
	\sum_{\kv'} 
	w_\alpha(\kv') 
	{\Delta_{\kv'}  \over 2 E_{\kv'}} 
	=
	\sum_\alpha A_\alpha w_\alpha(\kv) 	
	\,,
	\label{Deltaequation}
\eeq
where the new coefficients $A$ satisfy:
\beq
	A_\alpha =  	- 
	v_\alpha \sum_\beta 
	W_{\alpha \beta}[A] A_\beta
	\,,
	\label{Aeqs}
\eeq
and we have defined the matrix $W$, analogous of the matrix $B$ for 
the two-body problem:
\beq
	W_{\alpha \beta}[A] = 	
	\sum_{\kv} 
	{w_\alpha(\kv) w_\beta(\kv)  \over 2 E_{\kv}}
	\,.
\eeq
The $A$ dependence comes through $\Delta$ in $E$.
In matrix form Eq. \refe{Aeqs} reads  simply $(1+\Vc \,W)A=0$.
As before it is convenient to write the equation for $A$ in terms
of $\Vc^{-1}$, in order to perform easily the limit $U\rightarrow \infty$.
Eq. \refe{Aeqs} thus becomes:
\beq
	(\Vc^{-1}+W[A]) A = 0
	\,.
	\label{Aeq}
\eeq

In order to find $\Delta_\kv$ we need to solve the nonlinear equation
\refe{Aeq}, analogous of the linear equation \refe{selfequations} for
the two-body problem.
Let $\lambda_i$ and $v_\alpha^i$ be the eigenvalues and eigenvectors
of $\Vc^{-1}+W$. We project $A_\alpha= \sum_i c_i v^i_\alpha$ on 
the $v_\alpha^i$, so that 
\beq
	\sum_i \lambda_i[c] c_i v^i = 0
	\label{eigenform}
\eeq
$c_i$ is non vanishing only if $\lambda_i=0$. 
The corresponding coefficient $c = c_i$ is fixed by the equation $\lambda_i[c]=0$.
The form of the solution for $A$ is simply 
$A_\alpha = c v_\alpha^i$, and for 
$\Delta_\kv = c  \sum_\alpha v^i_\alpha w_\alpha(\kv)$.

Due to the symmetry of the crystal, the matrix $W$ 
is already in the form \refe{eigenform} for $\alpha=$ 2, 3, and 4.
The conditions $\lambda_i=0$ in these cases give the two equations
\beqa
	{1\over 2V} &=&  
	{1\over 2} \sum_{\kv} {
	[\cos(k_x a)-\cos(k_y a)]^2
	\over 2 E_\kv}
	\,,
	\label{DwaveSC}
	\\
	{1\over 2V} &=&  
	\sum_{\kv} {\sin^2(k_x a) \over 2 E_\kv}
	\label{PwaveSC}
	\,,
\eeqa
that correspond to the $d$- and $p$-wave solutions.
In the ``$s$'' subspace $\{w_0, w_1\}$ we are instead left with 
a 2 by 2 matrix:
\beq
	\Vc^{-1}+W[A] = 
	\left( \begin{array}{cc} 
	W_{00} & W_{01} \\
	W_{01} & -{1/2V}+W_{11} \\
	\end{array}
	\right)
	\label{vwmatrix} 
	\,.
\eeq
The condition of vanishing of the determinant gives:
\beq
	-{1\over 2V} = {W_{01}^2 \over W_{00}} -W_{11}  
	\,,
	\label{SwaveSC}
\eeq
When a solution is found the form of the gap function will be 
given by the corresponding eigenvector of \refe{vwmatrix}: 
$v = (W_{01}, -W_{00})$, thus 
\beq
	\Delta_\kv = 
	\Delta\left[
	1 - r {\cos k_x a +\cos k_y a \over \sqrt2}
	\right]
	\label{DeltaDef}
\eeq
where 
\beq
	r = W_{00}/W_{10}
	\label{requation}
	\,.
\eeq

The mixing of the two $w$'s is crucial in order to take into 
account correctly the hard-core repulsion. 
This was already clear from the two-body problem, where the exact 
solution gives a wave function that is a superposition
of $w_0$ and $w_1$. 
At the MF level we could, in principle, search for solutions with only
one of these two components different from zero:
such a procedure is wrong.
The simplest way to see that is to evaluate the anomalous correlation
function 
$
	\Fc(\xv-{\bf y})
	=
	\qav{c_{\uparrow}(\xv) c_{\downarrow}({\bf y}) }
$.
This function is the equivalent of the wavefunction for the two-body
problem: it gives the probability of finding two fermions on the same
site, thus for infinite $U$ it is bound to be zero at the origin.
We can verify that our solution correctly gives $\Fc(0)=0$:
\beqa
	\lefteqn{\Fc(0) 
	=
	\sum_\kv \Fc_\kv 
	= \sum_\kv {\Delta_\kv \over 2 E_\kv}}
	\nonumber \\
	&& \propto   \left[ 
		W_{01} \sum_\kv {1\over E_\kv} 
	    	-W_{00} \sum_\kv {w_1(\kv) \over E_\kv}\right]
	= 0 
	\label{Ffunction}
\eeqa
It is clear that if, out of the two order parameters, only one is present,
$\Fc(0)$ would be always non-vanishing. 
For $p$- and $d$-wave, the symmetry alone is sufficient to
guarantee the vanishing of the wavefunction.

The condition \refe{Ffunction} has important consequences on
the superconducting order. 
It is clear that in order to fulfill \refe{Ffunction} 
$\Delta_\kv$ must change sign in the Brillouin zone. 
In particular, since $w_1(\kv) = -t_\kv/(2\sqrt2 t)$, 
$\Delta_\kv$ is constant on surfaces of fixed energy:
\beq
	\Delta_\kv = \Delta(\xi_\kv)
	=
	\Delta[
	1+r 
	(\xi_\kv+\mu')/(2\sqrt2 t)
	]
	\,.
\eeq
Thus the line of nodes corresponds to a given energy $\xi_\kv$ instead
of given direction, as it would be in $p$- or $d$-wave
superconductivity.
The  energy for which $\Delta$ vanishes is 
$\xi_N=-{2\sqrt2 t/r}-\mu'$.
Other consequences of \refe{Ffunction} will be discussed in more
details in Section \ref{DeltaNodes}, after the solution of the
self-consistent equations.

Before proceeding, we find a simpler expression for the 
relevant $W$'s.
Adding and subtracting $\mu'$ in the numerator of $W_{10}$ we have:
\beq
	W_{10} 
	= 
	-{\mu'\over 2\sqrt2 \,t} W_{00}-
	{1\over 2\sqrt2\,t} \sum_\kv {\xi_\kv \over 2 E_\kv}
	\,.
\eeq
We use Eq. \refe{nequation} fixing $n$ to eliminate the 
second term. We obtain:
\beq
	2\mu' W_{00}+ 4\sqrt2\,t   W_{10} = n-1\,,
\eeq	
that is a generalization of \refE{ee1} for the two-body problem.
Similarly for $W_{11}$ we have:
\beq
	W_{11} = {\mu'^2\over 8t^2} W_{00} 
	+ { \mu'(1-n) \over 8\, t^2}  
	+ b
\eeq
where 
\beq
	b 
	= {1\over 16 t^2} \sum_{\kv} {\xi_\kv^2 \over E_\kv}
	= - {\mu' \over 16 t^2}
	+{1\over 16 t^2} \sum_\kv {\xi_\kv\over E_\kv}
	(\xi_\kv-E_\kv)
\eeq

In conclusion we obtain the following set of equations:
\beqa
	{1\over V} 
	&=& 
	2b(\mu',\Delta,r) -  {(n-1)^2 \over 16 t^2 W_{00}(\mu',\Delta,r)}	
	\label{terza}
	\\
	{1 \over r} 
	&=& 
	-{\mu' \over 2\sqrt 2\,t} + 
	{n-1 \over  4 \sqrt2\,t W_{00}(\mu', \Delta,r)}
	\label{quarta}	
\eeqa
This form is more convenient for the following discussion since 
$b$ remains finite when $\Delta\rightarrow 0$.

Equations \refe{terza} and \refe{quarta}, together with \refE{nequation},
form the complete set of equations for the three unknowns $\{\mu',\Delta,r\}$
(the gap is defined in \refE{DeltaDef}).
This will be our starting point to discuss the physics of the 
$s$-wave superconducting state.

\subsection{Threshold line for $s$-wave superconductivity}

\begin{figure}
\psfig{file=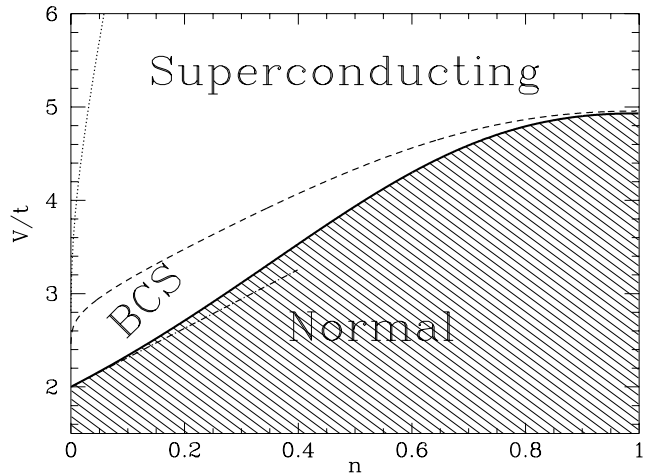,width=9cm,angle=-90}
\caption{
Boundary between the $s$-wave superconductor and a normal 
phase. 
The dashed line is the line where $\Delta/(\mu'+D)=.1$,
thus it gives an indication of the BCS region. 
The dotted line is the line where the chemical potential it 
exactly at the band edge: $\mu'=-D$, and can be considered 
the beginning of the Bose region.
The dot-dashed line in the dashed region is the 
analytical expression of \refE{Vana} valid for small $n$.
}
\label{FigVt}
\end{figure}
We have seen that in the two-body problem a pair is bound only if $V$
exceeds a threshold potential $V^s_t$.
For small densities it is clear that for $V>V_t^s$ the 
system is superconducting and in the Bose limit, 
{\em i.e.} with a coherence length much smaller that the 
average interparticle distance.
As shown rigorously by Randeria {\em et al.} \cite{Randeria2D},
in the low density limit and in 2D, the presence of a bound state
in the two-body problem is a necessary and sufficient condition
for $s$-wave superconducting instability to take place.
We thus expect that for $V<V_t^s$ and $n\rightarrow 0$ the 
system is not superconducting.
A separation line, starting at $V_t^s$ for $n\rightarrow 0$, as
function of $n$, separates the $s$-wave superconducting region and the
normal behavior.
The equation for this line of second-order transition can be 
found by setting $\Delta=0$ in \refE{terza} 
$n$ being fixed by the chemical potential $\mu'$
($-4t<\mu'< 4t \equiv D$).
We obtain:
\beq
	{1\over V} = -{\mu'\over 8 t^2} 
	+{1\over 8 t^2}
	\sum_\kv\left[
		|\xi_\kv|-\xi_\kv
	\right]
	\,. \label{lineacritica}
\eeq
When $\Delta=0$, $\mu'+D=\epsilon_F^o$ is the Fermi energy of an 
ideal gas measured from the band edge.
For small doping $\epsilon^o_F= 2\pi t n$, and since for small $n$ the
integral in \refe{lineacritica} is quadratic in $\epsilon_F$
we can  neglect it and find a simple analytic expression for the 
threshold curve:
\beq
	{V^s_t\over t} = 2 + \pi n + O(n^2)
	\label{Vana}
	\,.
\eeq
The complete curve is shown in Fig.~\ref{FigVt}. 
The critical line increases smoothly from $V/t=2$ for $n=0$ to 
$V/t \approx 4.93$ at half-filling ($n=1$). 
This can be easily understood by the fact that the particles feel more
and more the presence of the hard-core repulsion.
\begin{figure}
\psfig{file=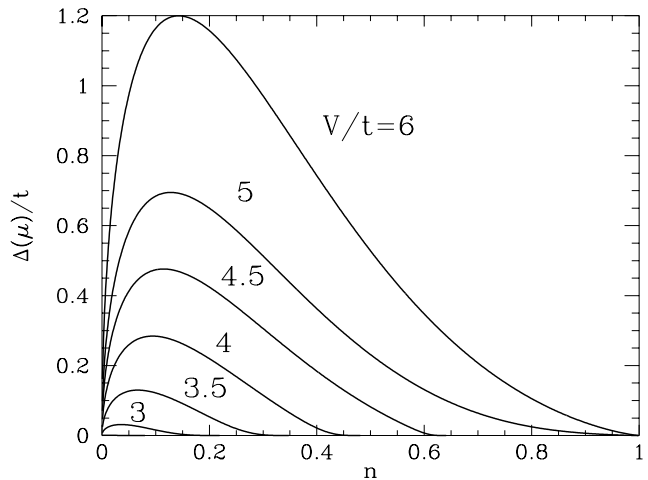,width=9cm,angle=-90}
\caption{
Gap function at the Fermi energy $\Delta(\mu')/t$ as a function
of the density $n$, and for different values of the coupling $V$.
Starting with the upper curve $V/t$ is: 6, 5, 4.5, 4, 3.5, 3, and 2.5.
One can easily see the square-root behavior for small $n$,
typical of a Bose behavior. For $ V<4.93$, near the critical 
point value of the density a BCS exponential decay is evident.
For $V>4.93$ and at half-filling, the gap function $\Delta(\mu')$
vanishes on the Fermi surface remaining finite elsewhere.
Since at half-filling the model fails, we do not discuss further this
behavior.  }
\label{DeltaFig}
\end{figure}

\subsection{Near the threshold: BCS limit}

\begin{figure}
\psfig{file=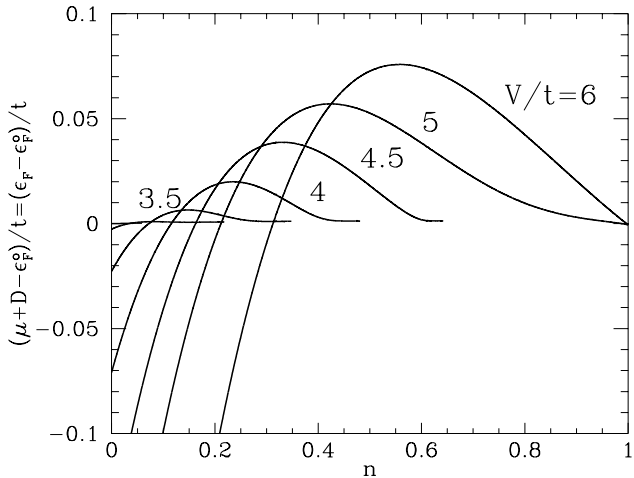,width=9cm,angle=-90}
\caption{
The chemical potential $\mu'$ measured from the bottom of the 
band $-D$, minus the free gas Fermi energy $\epsilon_F^o$.
The curve are at fixed $V/t$ which takes the values: 6, 5, 4.5, 4, 3.5, 3, 
and 2.5 from the top to the bottom curve.
The comparison is particularly interesting near the critical value of
$n$.  
There the behavior is BCS and the system is only slightly
different from the free gas, since $\epsilon_F \approx \epsilon_F^o$
and $\Delta\ll\epsilon_F$.
For $V$ larger that 4.93 no critical density exists, and the system 
never become strictly BCS.
}
\label{muFig}
\end{figure}

Near the transition line the system behaves
like a BCS superconductor for weak coupling. 
In particular, the superconducting gap is much smaller 
than $\epsilon^o_F$, the Fermi energy of the free gas,
and $\mu'$ remains $-D+\epsilon_F^o$. 
In order to prove this fact we assume that these properties
hold, and we evaluate \refE{terza}. 
$W_{00}$ has the usual  $\ln \Delta$ divergent behavior
and it can be evaluated in leading order by assuming 
a constant density of states and constant $\Delta_\kv$ near
the Fermi surface ($t_\kv=\mu'$).
It is convenient to write the equation in terms of $V-V^s_t(n)$,
by subtracting the critical line equation. 
In this way we obtain: 
\beq
	\Delta(\mu') \approx 2\sqrt{D^2-\mu'^2} \exp 
	\left\{	
	-{ (1-n)^2 V V_t(n) \over 16 t^2[V-V_t(n)]\rho(\mu')}
	\right\}
	\,,
\eeq
where $\rho(\mu')$ is the density of states per spin evaluated
at $\mu'$.
Near the threshold $\Delta/\epsilon_F^o \ll 1$ and thus the
superconductor is of BCS type.
Such a regime persists as long as $\Delta(\mu')/(\mu'+D)$ is small.
The dashed line in \refF{FigVt} indicates the values of $n$ and $V$
for which $\Delta(\mu')/(\mu'+D)=0.1$.
The ``BCS region'' is restricted between that line and the
critical line $V_t^s$ -- a narrow region compared to the remainder of the
phase diagram.
The behavior of the chemical potential is shown in Fig.~\ref{muFig}.

\subsection{Bose limit and intermediate regime}

%
%
%
%
%

%
%
%
%
\begin{figure}[tb]
\psfig{file=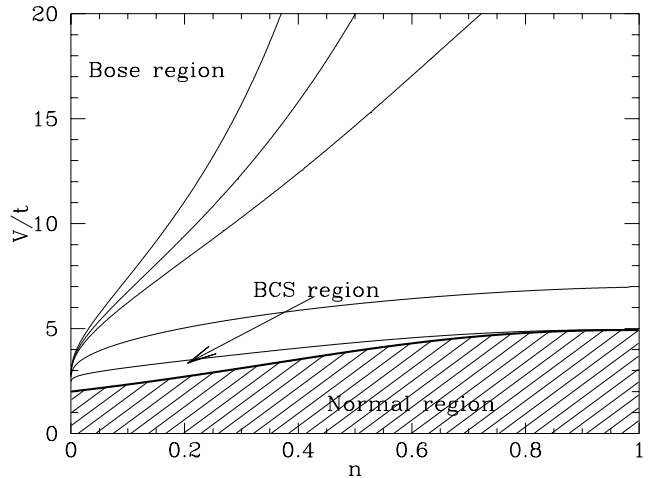,width=9cm,angle=-90}
\caption{
Phase diagram of the model for $n$ and $V/t$, $U=+\infty$. 
In the shaded region $s$-wave superconducting is not stable. 
In the rest of the diagram we find that $s$-wave superconductivity
smoothly crosses from the BCS regime, near 
the critical line (the border of the shaded region), to
the Bose regime, near the $n=0$ line.
In order to characterize better this crossover we report the 
lines of fixed ratio $\Delta(\mu')/(\mu'+D)$. 
From the upper line to the critical line the ratio is 
$\infty$, 10, 5, 1, and .1.
Surprisingly the BCS region is narrow, and the genuine 
crossover region covers most of the diagram.
}
\label{ratioFig}
\end{figure}

As discussed above, for $V>V_t^s$ and small density we 
inevitably reach the Bose limit. 
This is confirmed by the mean field equations. 
Indeed for $n\rightarrow 0$ the quantity $\Delta/|\mu'+D|$ vanishes and 
$\mu'<-D$.
One can thus verify that the \refE{terza} reduces to 
\refE{finalswave} for the bound state energy of the two-body problem
with $\mu'=\omega_B/2$.
At the same time $\Delta$ is fixed by \refE{nequation} that
gives $\Delta \sim t \sqrt n$. 
This behavior of $\Delta$ can be seen in Fig.~\ref{DeltaFig}
for small $n$.
For intermediate couplings and densities the numerical solution 
crosses over between the BCS and Bose limits described above.
In Fig.~\ref{ratioFig} we plot the lines of constant
$\Delta(\mu')/(\mu'+D)$.
We have seen that this ratio characterizes both the BCS (small
positive value) and the Bose limit (small negative value).
The phase diagram for the crossover from BCS to Bose condensation
in the purely attractive case was given in Ref. \onlinecite{andrenacci}
for both $s$-wave  (local attraction) and $d$-wave (nearest neighbour 
attraction).
A comparison of the results shows that the effect of the
repulsion is to eliminate a large portion of the $s$-wave BCS region, which
becomes normal. The $d$-wave case will be discussed in 
Section \ref{dwave}.

All these curves are plotted in the range of densities (0,1): 
it should be stressed that the have no real meaning near half
filling since the mean field approach misses the insulating 
state. 
We expect that the approximation remains valid as far as $z_kv$, the
weight of the quasiparticle peak, remains of the order of 1.
We also remember that in the corresponding Kondo problem
the limit $n=1$, corresponds to no particles at all.

\subsection{Nodes of $\Delta(\xi)$}

\label{DeltaNodes}

We have seen above that $\Delta$ depends only on $\xi$ and that 
it vanishes for $\xi=\xi_N$. 
As a consequence we can write $\Delta(\xi)$ as follows:
\beq
	\Delta(\xi) = { \Delta  r\over 2\sqrt 2 \, t} [\xi-\xi_N] 
\eeq
where, substituting \refE{quarta} in the expression for $\xi_N$ we have:
\beq
	\xi_N = { 1-n \over 2  W_{00}(\mu',\Delta,r)}
	\,.
\eeq
For small value of $\Delta$,  $\xi_N \sim -t/\ln(\Delta/t)$, thus 
it vanishes logarithmically in $\Delta$.
This means that approaching the threshold line $V^s_t$ the lines of
nodes approach the Fermi surface.
It turns out that keeping it at a logarithmically small distance (in
$\Delta$) suffices to fulfill the hard-core condition \refe{Ffunction}.

\begin{figure}[tb]
\psfig{file=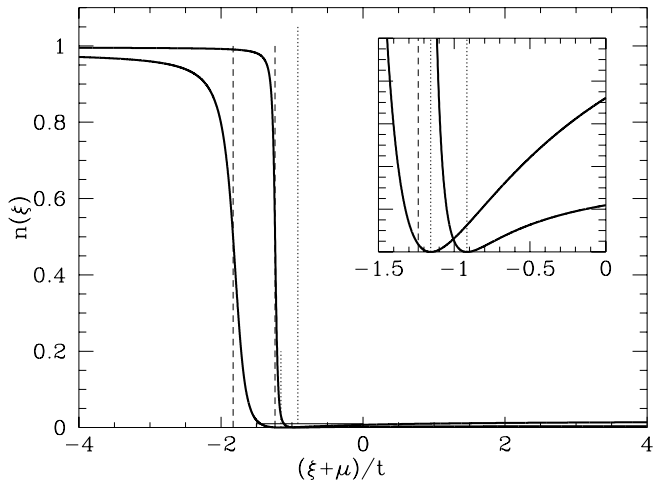,width=9cm,angle=-90}
\caption{
Occupation number $n_\kv=n(\xi_\kv)$ for $V/t=4.5$ and
two different values of $n$: 
$n=0.4$ (left curve) and $n=0.642$ (right curve).
In the inset the region where $n_\kv$  and $\Delta_\kv$ 
vanishes is shown enlarged. 
The dashed line indicates the value of the chemical potential, and 
the dotted one $\xi_N$.
}
\label{Fignk}
\end{figure}

The presence of a line of nodes near the Fermi surface modifies the
usual shape of the occupation number distribution $n_\kv$.
Specifically, it introduces a particle-hole asymmetry for
$|\xi|\approx \xi_N$.
For large values of the energy the $n$ distribution is 
anomalous. 
Usually, far from the Fermi surface, $n_\kv$ is either 0 or 1,
depending on the sign of $\xi$.
In contrast our $\Delta(\xi)$ is $ \sim \xi$ when 
$|\xi|$ is large with respect to both $\Delta$ and $\xi_N$,
leading to 
\beq
	n(\xi)
	\approx 
	{1\over 2} 
	\left[ 
	1-\sgn(\xi) {1\over \sqrt{1+\Delta^2 r^2/8t^2}}
	\right]
	\,.
\eeq 
This behavior can be recognized in Figure \ref{Fignk}
where the form of $n(\xi)$ is shown
for $V/t=4.5$ and for two different values of $n$.
Inspection of that figure shows that $n(\xi)$ remains
symmetrical around the chemical potential only for $|\xi|\ll
\Delta(\mu')$.
The asymmetry for $|\xi|> \xi_N$ is clearly apparent in the inset.

\begin{figure}[htb]
\psfig{file=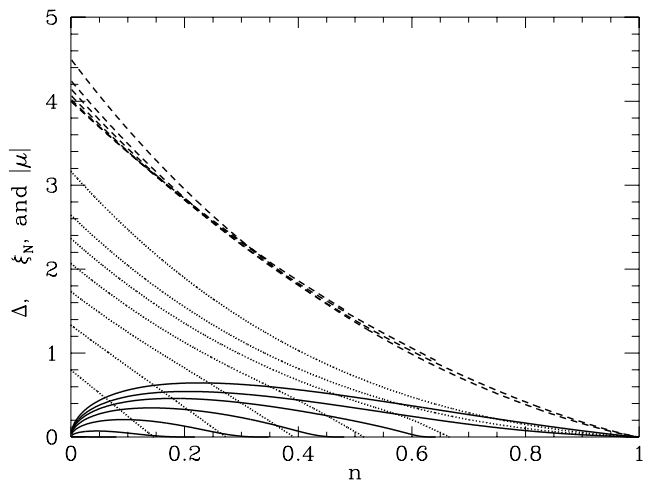,width=9cm,angle=-90}
\caption{
 The three relevant energy scales of the problem: (dashed lines)
$|\mu|$, (dotted lines) $\xi_N$, and (full line )$\Delta(\mu)$ are
plotted for the same values of $V$ of Fig.\ref{DeltaFig}.
As expected, $\xi_N$ vanishes linearly in $n-n_t$, while 
$\Delta$ vanishes exponentially in $1/(n-n_t)$. 
}
\label{EnergyScales}
\end{figure}

The aforementioned asymmetry explains the anomalous behavior of the 
chemical potential in the superconducting state 
seen in Fig.~\ref{muFig}.
The chemical potential is normally shifted down with respect to the
free gas value!  
The BCS distribution function is particle-hole
symmetric, and states are ``missing'' below the lower band edge:
$\mu$ must go down in order to retain the same density.
In contrast here the lower spectral density near $\xi=\xi_N$ leads to
an increase of $\mu$ in a narrow range where $\Delta$ and $\xi_N$ are
comparable, as shown in Figures \ref{muFig} and \ref{EnergyScales}.

\section{$d$-wave superconductivity}
\label{dwave}

We consider now the possibility of $d$-wave superconductivity.
By symmetry, the hard-core part of the interaction is not seen
by the two-body wave function, thus the scattering states 
should always feel an attraction at finite energy. 
This is enough to induce BCS superconductivity at any finite density
and for any value of the coupling constant $V$.
In contrast the Bose limit can be reached only if a bound state 
is present in the two body problem, $V > V_t^d = 7.35\,t$.
Remember that in this limit $s$-wave always wins,
we thus restrict to $V<V_t^d$.

In the opposite limit, for small values of the coupling 
$V \ll V_t^s$, the situation is again clear: we expect a standard BCS
$d$-wave superconductors, since there is no $s$-wave solution there.
(The presence of a small hopping at nearest neighbors is discussed in
the next section and it will not change the final outcome.)

The situation is much more interesting at intermediate couplings,
$V_t^s < V < V_t^d$. 
In this region there is competition between the two possible
superconducting orders: the lowest in energy will be the actual ground
state.
We analyze qualitatively this competition as a function of the density $n$
for a fixed value of the coupling $V$.
For small density ($n\rightarrow 0$) the $d$-wave order parameter
has a BCS-like form with an effective interaction 
$V_{eff} = V\, \qav{w_2^2}_{FS}$ where  
$\qav{w_2^2}_{FS}$ is the average over the Fermi surface of the 
angular factor. 
Since $w_2(\kv) \sim (k_x^2-k_y^2)$, $V_{eff}$, which is of order $w_2^2$,
is reduced by a factor $n^2$. 
The gap $\Delta_\kv$ is exponentially small and the resulting energy
gain vanishes extremely rapidly with $n$ since $\delta E \approx
-\rho(\mu) \qav{\Delta^2_\kv}_{FS}/2$.

The $s$-wave solution, instead, is in the Bose limit, thus its
energy gain is simply the number of bosons multiplied by the 
bound state energy: $\delta E = n \omega_B/2$. 
For $n\rightarrow 0$ the $s$-wave is always preferred to the $d$-wave.

The opposite holds near the threshold for $s$-superconductivity:
there $d$-wave still lowers energy, and is thus preferred.
We expect a first order transition in between. The 
question is ``where''?
Will the BCS $s$-region be completely sweeped out by the $d$-wave BCS
state?
\begin{figure}[tb]
\psfig{file=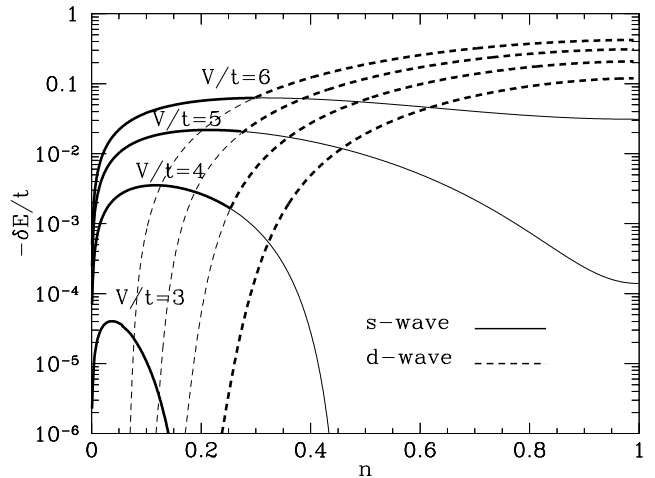,width=9cm,angle=-90}
\caption{ Ground state energy gain $|\delta E|$ for $s$-(full lines)
and $d$-wave (dashed lines) superconducting state as a function of the
density for different values of $V$ (indicated in the figure).
The realized state is emphasized with a thicker line.
}
\label{GSE}
\end{figure}

In order to answer this question we need an explicit calculation 
of the ground state energy gain for the superconducting state:
\beq
	\delta E = 
	2 \sum_\kv t_\kv n_\kv 
	-2 \sum_{\kv<\kv_F} t_\kv 
	-\sum_\kv \Delta_\kv \Fc_\kv
	\label{deltaE}
	\, .
\eeq
The last term in \refe{deltaE} is $-\Delta^2_d/(2V)$ for the $d$-wave
symmetry, and $\Delta^2 r^2/(2V)$ for $s$-wave.
$\Delta_d$ is obtained by solving \refE{Dwave} and \refE{nequation}
with $\Delta_\kv = \Delta_d\, w_2(\kv)$.
$\Delta$ and $r$ are the quantities introduced in \refe{DeltaDef} and
discussed in the previous Section.
Numerical results are shown in Fig.~\ref{GSE}.
One can clearly see the crossing of the two energy values indicating
the first order transition.
For small density, as expected, the $s$-wave state is always lower in
energy (left side of the diagram).
Note however that the value of the energy for which the transition
takes place depends strongly on the value of $V$ considered.  
We can distinguish two cases:

({\em i}) $V$ slightly above $V_t^s(n=0)$ the $s$-wave
superconductivity spans the whole crossover from the Bose limit 
(for $n\rightarrow 0$) to the BCS one.
The first order transition to the $d$-phase occurs deep in 
the $s$-BCS region.
Note that this happens at tiny values of $\delta E$
(for instance for $V=2.5 \,t$,  $\delta E_c/t < 10^{-10}$).
Between the $s$- and the $d$-wave superconducting phases there exists
a region that for all practical purposes is normal.
Increasing the density at fixed $V$ one thus find that 
superconductivity is reentrant.

({\em ii}) At larger values of $V$, the $s$- and $d$-state energies
cross for a value of the energy gain reasonably large (of the same
order of the maximum value achieved along the $s$-wave evolution).
Thus a nearly monotonic behavior of $\delta E$ results (cf. the line
$V=6\,t$ ).

In conclusion we found that the $d$-wave competes with the 
$s$-wave only for large values of $V$. When $V$ is near the 
threshold a quasi-normal region between the $s$-wave and the $d$-wave 
can survive, and the BCS-Bose crossover in $s$-wave can take 
place fully.

\section{Next neighbor hopping and $d$-wave $s$-wave competition}
\label{tprimo}

In this Section we consider the effect of the inclusion of a 
nearest-neighbors hopping integral $t'$.
The effect of this term on the superconducting critical temperature
has been investigated in Ref.~\cite{Ranninger} and in more details in
Ref.~\cite{BastideLacroix}.
Here we consider ground state properties and in agreement with
Refs.~\cite{Ranninger,BastideLacroix} we find that the introduction of 
$t'$ strongly modifies the $s$-wave phase diagram.
As a matter of fact, the threshold line $V_t^s$ is
washed out by an arbitrary small $t'$.
A transparent proof of that has been given in Ref.~\cite{BastideLacroix}.
We reproduce here their arguments for zero temperature.
Equation \refe{SwaveSC} for $\Delta$ can be rewritten as follows:
\beq
	{1\over 2V W_{00}} = \qav{w_1^2 } - \qav{w_1}^2
	=\qav{(w_1 - \qav{w_1})^2}
	\,,
	\label{SwaveSC2}
\eeq
where we have defined:
\beq
	\qav{g(\kv)} \equiv 
	\sum_\kv 
	{g(\kv) \over E_\kv} 
		/
	\sum_\kv 
	{1\over E_\kv} 
	\,.
\eeq	
When $\Delta$ is small, the weighted average over the Brillouin
zone is peaked on the Fermi surface.
If $w_1(\kv)$ has the same $\kv$ dependence of the kinetic energy, the
righthand side of \refe{SwaveSC2} vanishes at leading order in
$\Delta$, and the logarithmic divergence in $W_{00}$ will be
ineffective: superconductivity exists for $V$ larger than a threshold
$V_t$.
In contrast, when the spectrum $t_\kv$ is such that $w_1(\kv)$ 
varies on the Fermi surface, the right-hand side of 
\refe{SwaveSC2} is finite for $\Delta\rightarrow 0$ and 
superconductivity is possible however small $V$.

This is discussed extensively in \cite{BastideLacroix}. 
Here we want to point out two interrelated facts: 
({\em i}) the relation of superconductivity with 
the presence of a bound state in the
two-body problem.
({\em ii}) The unusual nature of the superconducting state 
created by a small $t'$ in the region of the phase diagram where
the system would be normal but for $t'$.

Randeria {\em et al.} \cite{Randeria2D} proved, within a
two-dimensional continuum model, that the existence of an $s$-wave
bound state for the two-body problem is a {\em necessary} and
{\em sufficient} condition for the existence of $s$-wave
superconductivity at low density.
In the model at hand the presence of a small $t'\neq 0$ 
makes superconductivity always possible while it maintains
a threshold for the two-body bound state.
That threshold stems from the fact that $w_1$ has no fluctuations
near the band edge -- hence no $\log \omega_B$ term in 
the pair propagator.
This result seems to contradict the prediction of
Ref.~\cite{Randeria2D}.

The solution of this inconsistency comes from the $\kv$ dependence 
of $\Delta$ at the Fermi surface. 
In fact, even if the order parameter is of $s$-wave symmetry, it
actually changes sign 8 times on the Fermi surface and 
it has a zero average on it.
This can be verified by solving the set of Eqs. \refe{DwaveSC} and
\refe{nequation} in power series of $t'$.
One finds the following expression for $\Delta_\kv$:
\beq
	\Delta_\kv 
	= 
	\Delta \, {t' \over \mu'} 
	\left[v(\kv)-\qav{v}_{FS}\right] 
	+ O({{t'}^2 / \mu'}) 
\eeq 
where $t_\kv = -2t\sqrt{2}\, w_1(\kv)+ t' v(\kv)$.
More generally, it is clear that for any perturbation $v(\kv)$ to the
nearest neighbor hopping, $\Delta_\kv$ will change sign on the Fermi
surface, and since it has $s$-wave symmetry this must happen at least
8 times. 

Due to the unusual behavior of $\Delta_\kv$  the results
of Randeria {\em et al.} do not
apply to this $s$-wave state which is, in practice, close
to an angular momentum $l = 4$ state
(which is isotropic as regards cubic symmetry.)

Hence  $s$-wave superconductivity survives the hard core
for arbitrary small values of the attraction.
But since the superconducting state is close to 
$l=4$, the resulting value of $\Delta$ (or $T_c$) is 
extremely small, even compared to the $d$-wave solution.
We checked explicitly this fact for our model. 
We considered values of $t'< t/2$, since at this value the Fermi
surface becomes double sheeted.
We calculated the effective $s$-wave coupling 
\beq
	V_{eff}^s/V 
	= 
	\qav{w_1^2} - \qav{w_1}^2
	\sim 
	\left[
	\qav{v^2} - \qav{v}^2
	\right] t'^2/(8t^2)
\eeq
as a function of the Fermi energy and of the coupling $t'$.  
We compared $V_{eff}^s$ with the effective $d$-wave coupling:
$V_{eff}^d/V = \qav{w_2^2}$.
In all cases we found that $V_{eff}^d>V_{eff}^s$.

We thus conclude that the introduction of a $t'<t/2$, or of other 
similar perturbations, does not change qualitatively the above discussion,
since $d$-wave superconductivity always hides this anomalous $s$-wave
state.

\begin{figure}[htb]
\psfig{file=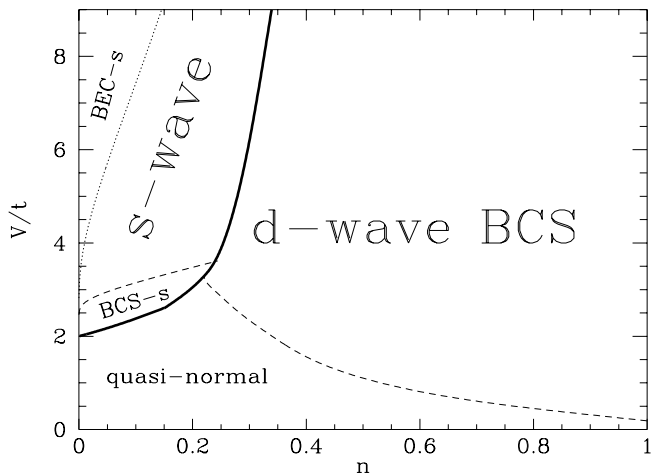,width=9cm,angle=-90}
\caption{ 
Phase diagram considering the competition between $s$- and $d$-wave.
The thick line corresponds to the first-order phase transition
between the $s$- and $d$-wave superconducting order. 
On the $s$-side we indicate the BCS and Bose condensation regions.
On the $d$-side the region where $\delta E/t<10^{-6}$ is labeled 
as ``quasi-normal''. 
}
\label{FinalPD}
\end{figure}

\section{Conclusions}
\label{conclusions}

We have studied the appearance of superconductivity 
on a 2D lattice in presence of a hard-core repulsion and 
of a nearest neighbor attraction.
We constructed the mean field solution and compared it 
to the two-body problem in vacuum.
Even if the procedure is restricted to mean field, it 
correctly forbids the double occupancy of the same 
site of the superconducting wavefunction.
The main results are the following:
(i) the $s$-wave solution is suppressed for small values of the coupling
at any density.
The introduction of additional small hopping integrals cannot change
this, since the $d$-wave solution is always preferred.
(ii) the $d$-wave solution is always possible, but the actual
value of the energy gain becomes extremely small at low density.
Thus for small $V$ and $n$ the critical temperatures are tiny:
in practice at any temperature the system is normal.
(iii) For $V$ larger than the threshold for the $s$ two body-bound state
the system exhibits a crossover from the
Bose-Einstein condensation of fermions pairs to a  BCS behavior
as the density is increased.
A first order transition to the $d$-wave
state occurs.
The situation is summarized in the phase diagram of Fig.~\ref{FinalPD}.

When only nearest neighbor attraction is present the competition between 
$d$-wave and $s$-wave at $T_c$ has been considered recently by
Wallington and Annett \cite{Wallington}. 
They found that for any coupling the $s$-wave has the lowest critical
temperature, but for a small region near half-filling, where the 
van Hove singularity of density of states weighted with the angular
factors stabilizes the $d$-wave phase.
The presence of the hard-core repulsion allows $d$-wave superconductivity
to appear on a much larger portion of the phase diagram.
The correct treatment of the repulsion is thus crucial to 
study the competition between $s$- and $d$-wave.

The technical procedure developed can be applied to
the Kondo-lattice problem sketched in the introduction.
However, one has to keep in mind that in that problem
superconductivity and Kondo hybridization should be considered on the
same footing, since each of the two order parameter reacts on the 
other. 
We will not try thus to apply the above result {\em directly} to 
the Kondo lattice model: we leave the detailed solution of 
the superconductivity-Kondo hybridization in presence of 
the hard-core repulsion for future work \cite{PistolesiNozieres}.


\begin{thebibliography}{99}
\bibitem{Qdep}
We consider only the case $\Qv=0$, however, since $ V_{\kv+\Qv/2,
\qv+\Qv/2} = V_{\kv, \qv}$ the same calculation gives the complete
spectrum with the simple substitution $2 t_\qv \rightarrow
t_{\qv+\Qv/2}+t_{\qv-\Qv/2}$.
\end{thebibliography}
\end{document}